\documentstyle[12pt]{article}

\newlength{\extraspace}
\newlength{\extraspaces}
\newcommand{\be}{\begin{equation}}
\newcommand{\ee}{\end{equation}}
\newcommand{\bea}{\begin{eqnarray}}
\newcommand{\nn}{\nonumber}
\newcommand{\eea}{\end{eqnarray}}

\newcommand{\nk}{\noindent}

\newcommand{\eqn}[1]{(\ref{#1})}
\def\lsim{\mathrel{\rlap {\raise.5ex\hbox{$ < $}}
{\lower.5ex\hbox{$\sim$}}}}

\newcommand{\pr}{\paragraph{}}
\def\gappeq{\mathrel{\rlap {\raise.5ex\hbox{$>$}}
{\lower.5ex\hbox{$\sim$}}}}
\def\lappeq{\mathrel{\rlap{\raise.5ex\hbox{$<$}}
{\lower.5ex\hbox{$\sim$}}}}

\begin{document}

\thispagestyle{empty}

\begin{flushright}
{\sc OUTP} -96 - 66--P\\
DSF 53/96\\
hep-th/9611040\\
November 1996
\end{flushright}
\vspace{.3cm}

\begin{center}
{\large {\sc {\ Quantum Phase Space from String Solitons } }} \\[20pt]

{\sc Fedele Lizzi}\footnote{
On leave from Universit\`a di Napoli Federico II, Dept.\ of Physical
Sciences, Pad.\ 19 Mostra d'Oltremare, 80125 Naples, Italy.\\
e-mail: lizzi@na.infn.it} and {\sc Nick E.
Mavromatos}\footnote{
P.P.A.R.C. Advanced Fellow.\\
e-mail: mavroman@thphys.ox.ac.uk}\\[2mm]
{\sl University of Oxford, 
Theoretical Physics, \\[2mm]
1 Keble Road,
Oxford, OX1 3NP, UK.} \\[15mm]

{\sc Abstract}
\end{center}
In this paper we view the $\sigma$-model
couplings of appropriate vertex operators describing the interaction 
of string matter with a certain type of string solitons
(0-branes) as the
quantum phase space of a point particle.
The $\sigma$-model is {\em slightly non critical}, and
therefore one should dress it with a Liouville mode.  
Quantization is
achieved by summing over world-sheet genera (in the pinched
approximation).  
To leading order
in the coupling constant expansion, the quantization reproduces the
usual quantum mechanical commutator.  We attempt to go beyond leading
order, and we reproduce the generalized string uncertainty principle.

\noindent

\vfill

\newpage \pagestyle{plain} \setcounter{page}{1} \stepcounter{subsection}

\section{Introduction.}

The concepts of space and time and its quantization are still far from being
understood. One of the most promising hopes for a 
reconciliation of quantum Gravity with quantum mechanics seems,
at present, available within the framework of string theory. The
underlying two-dimensional conformal field theory structure of strings
provide a field-theoretic framework for a consistent formulation of
target-space quantum gravity as an effective theory of strings below the
Planck (or String) scale.

One of the most fruitful ways to look at string theories, in turn, is to
see them as two dimensional $\sigma$-models, that is field theories on a two
dimensional 'space--time', the world--sheet, on which some fields are
defined, those fields are the coordinates of a `target' space time on which
the string moves. 

Admittedly, this world-sheet description of string theories may not be
sufficient for a complete non-perturbative description of quantum
space time effects. Target-space instantons in certain string
backgrounds may not have their analogues as world-sheet elevated
objects of a conformal field theory \cite{inst}. However, for most
purposes a world-sheet conformal field theory may prove sufficient.

In this paper we would like to go a step further. We will consider the
target space of the fields defined on the world sheet as the {\em
scaffolding} necessary to build solitons. We will then consider as 
coordinates of our phase space 
the $\sigma$-model couplings of the vertex operators of
these solitons. This idea is reminiscent of the world-sheets for
world-sheets idea of M.B.~Green \cite{green}, in which the target
space of one 
(two-target-space-dimensional)
string theory is considered as the world-sheet of the
`next' theory.  
Here it is rather (critical-dimension) 
solitonic stringy space times for
phase-spaces.  
What we will try to do, is to quantize the coordinates
of this {\em theory} space. This is a space with a very rich structure,
as the work of Zamolodchikov on the renormalization group flow has
shown \cite{zam}.  A quantization of such a space is achieved by
summing over topologies on the world sheet \cite{emninfl,emnd}.

Our action will be the usual $\sigma$-model action for strings with
the addition of terms describing the boundary conditions appropriate
for a soliton (D-brane) background \cite{dbranes,callan}. This action 
has infinities coming
from the coupling of string matter to the soliton, corresponding (from
the scaffolding space point of view) to the recoil of the D-brane due
to scattering \cite{recoil,kogmav,periwal,kogwheat}. To cancel these
infinities one has to add these recoil operators to the action. It
turns out \cite{kogmav,kogwheat} that the recoil operators, being
solutions of a degenerate hypergeometric equation, come in {\em
logarithmic} pairs \cite{gurarie,tsvelik}. It is the couplings of this
pair of operators which play the role of position and momentum in the
approach we propose here. The appearance of a {\em phase space} is due
to the mixing of these logarithmic operators.

Quantization is then achieved by observing that, due to
renormalization, the theory is {\em slightly non critical}
\cite{aben,emn}. To restore criticality at the scaffolder 
level\footnote{By {scaffolder point of view} we mean the (usual) point of
view in which the coordinates of space time are the fields $X$ defined on
the two dimensional world sheet. },  one adds the  Liouville
field.
In fact, a finite scale transformation on the world-sheet, is a Galilean
transformation in our phase space. Thus one could also see the zero
mode of the Liouville field as the evolution parameter
\cite{emn,kogan,kogwheat,diffusion}.
The renormalization group flow for the Liouville field satisfies the
so-called Helmholtz conditions, a set of conditions which, if
satisfied, ensure the possibility of a {\em canonical} quantization
\cite{hojman,emninfl}.
We verify that the couplings of the logarithmic operators of the
D-brane satisfy these conditions, and therefore are canonically
quantized in theory space.

The quantization is achieved by a summation over world-sheet genera,
albeit in the approximation of pinched world sheets
\cite{emninfl,emnd,recoil,kogmav,schm}.  To leading order in the $\sigma$-model
coupling constant expansion, the canonical quantization reproduces the
usual quantum phase space with position and momenta having a constant
commutator.  We have however attempted to go beyond the leading
approximation, by incorporating stringy effects. We find corrections
to the commutator which reproduce the generalized string uncertainty
principle \cite{ven}. Let us however stress that our approach is
different. There, the case was made that trivial world-sheet topologies
gave the dominant contribution, due to some sort of `string asymptotic
freedom'. In our approach fluctuating topologies are crucial.

In this paper we are not aiming at giving a rigorous account of the
quantization of solitons in string theory. Instead, we would like to
suggest new ways of using string theories, and in particular the
non-critical ones, to explore the connections between space and
quantization. In this respect, we would like to mention 
that our use of Liouville dynamics will be physically different 
from that of ref. \cite{emn,emnd}. There, the Liouville 
field was identified with the scaffolder time $X^0$, which 
is not the case in this work. 

The structure of the article is as follows. In section 2 we review
briefly the $\sigma$-model formalism describing interaction of
D-branes with string matter, and introduce the logarithmic operators.
In section 3 we discuss how the space of the coupling constants
pertaining to these operators may be viewed as a quantum phase space.
In section 4 we present 
a $\sigma$-model path-integral approach to quantization,
summing over pinched world-sheet topologies. In section 5 we derive
the phase space commutators and reproduce the generalized 
uncertainty principle. 
In section 6 we collect some concluding
thoughts. Finally, some technical details related to logarithmic
operators are collected in the appendix.

\section{D-branes in $\sigma$-model Formalism\label{dbranes}}

Dirichlet-Branes (or D-branes) are solitonic states in open string
theories which appear if one considers
$10-(p+1)$ of
the (critical-dimension) (super)string 
coordinates to satisfy Dirichlet (rather than Neumann) boundary
conditions at the world-sheet boundary \cite{dbranes}. 
{}From the point of view of 
the scaffolding
space time, then, there are the zero modes of some 
target-space (collective) coordinates which do not
change with the world sheet coordinate $\tau$. There is, therefore, an object
which can be interpreted as a soliton at rest.

Such structures encode enormous information about non-perturbative
string symmetries, such as duality, connections among different string
theories etc.
In this section we review the relevant aspects of the D-brane
formalism in the $\sigma$-model approach. Far from being complete in
this discussion, we refer the interested reader to the very rich
literature on the subject, for example \cite{dbranes}.

In this paper we will consider only the case in which $p=0$, that is the
case of the so called $0$-brane. In this case, seen from the scaffolding
point of view, the soliton is zero dimensional and appears as a particle at
rest (with an opportune choice of reference frame).

The particular boundary conditions describing the D-brane can be given
a $\sigma$-model lagrangian formulation by considering a 
world-sheet boundary operator 
(see \cite{callan} and references therein):
\begin{equation}
{V}_D = \int _{\partial \Sigma} x_i \partial _n X^i + v_i X^0
\partial_n X^i  \label{dbraneop}
\end{equation}
where $n$ denotes the normal derivative on the boundary of the world
sheet $\partial \Sigma$, which at tree level is assumed to have the 
topology of a
disk of size $L$; $X^i~,i=1,\dots 9$ denote the collective excitations of
the brane satisfying Dirichlet boundary conditions on the world-sheet
boundary, while $X^0$ satisfies  standard Neumann boundary conditions, 
\begin{equation}
X^i ({\rm boundary}) = 0,~i=1,\dots, 9 \qquad \partial _n X^0 ({\rm boundary})
 = 0  \label{fiveb}
\end{equation}

The two couplings $x_i$ and $v_i$ in (\ref{dbraneop}) have the meaning of
the initial position and velocity of the free soliton respectively. Here and
in the following we shall limit ourselves to the case of a flat world volume.
Of course the lower case $x$'s, which are, from a two dimensional
point of view, couplings should not be confused with the $X$'s which
instead are fields. 

\subsection{D-branes Coupled with String Matter}

We will now consider the interaction of the soliton with string matter, and in
particular, in a perturbative spirit, with the lowest lying states of the
string, tachyons in the bosonic case\footnote{
As well known, bosonic strings suffer from the usual tachyonic
instabilities. Supersymmetry gets rid of such states, and also
stabilizes the solitonic backgrounds. Since in what follows we are
interested in the bosonic sector of the theory, we will ignore the
fermionic sector altogether. Inclusion of this sector would probably
induce a spin structure in our phase space. }.

Let us therefore consider for the moment the following action: 
\begin{equation}
  S_\sigma = \int _{\Sigma} [\partial X^\mu {\overline \partial}X^\nu
\eta_{\mu\nu} + T(X)] + 
 \int _{\partial \Sigma} x_i \partial _n X^i + v_i X^0
\partial_n X^i  
\label{action}
\end{equation}
where $\eta _{\mu\nu}$ is a (critical-dimension) 
flat Minkowski space-time metric, and $T(X)$ denotes
a bulk (closed string) matter operator, which for the case of bosonic
strings is taken to be the tachyon. 

We immediately encounter an infinity in the anulus correction to the tachyon
propagator in the presence of a D-brane, which expresses 
quantum effects in the interaction of matter (tachyons)
with the soliton (D-brane) background. 
One starts by considering the scattering of a closed string
tachyon (matter) state off a $0$ brane. For this purpose, it is
sufficient to
consider the anulus amplitude of two 
closed-string vertex operators $V(k)$, which must be integrated over the
propagating open string. This computation may be performed using the
operator formalism, in which one evaluates ${\rm Tr}V(k_{1})\Delta
V(k_{2})\Delta $, with $\Delta ^{-1}\equiv L_{0}-1$, where $L_{0}$ is the
Virasoro operator.

The part of this computation that is relevant for our purposes is the one 
due
to the world-sheet zero modes~\cite{recoil,kogmav,periwal}. Writing $\Delta
\equiv \int_{0}^{1}dq q^{L_{0}-2}$, $L_{0}=2p^{2}+N$, where $N$ is the string
level number, and picking out the $N=0$ part, we find the following
contribution to the anulus amplitude: 
\begin{equation}
{\cal A}=\int
dp<p|\exp(-ik_{1,i}^{0}X^{i,0})q_{1}^{-2(p^{0})^{2}}\exp(-ik_{2,i}^{0}X^{i,0})q_{2}^{-2(p^{0})^{2}}|p>
\label{anulus}
\end{equation}
where the integral is over momentum eigenstates of the Virasoro 
operator $L_0$~\footnote{Here one assumes diagonalizability
of $L_0$ in the space of states, which 
is justified in a leading-divergence approximation \cite{kogwheat,diffusion}. 
In the case of D-branes 
there are subleading divergences
coming from non-diagonalizable parts of $L_0$. Such 
non-diagonalizability is a generic
feature of a pair of 
logarithmic operators~\cite{tsvelik,kogmav,kogwheat}, 
leading to non-trivial mixing.
In the above analysis we determine one member of the 
pair, namely the one associated with leading divergences.
Later, we determine the other member by applying generic 
conformal field theory analysis \cite{kogwheat}.}, the 
superscript $0$ denotes time components in the scaffolder 
$\sigma$-model, and $q$ is the modular parameter of the anulus. 
The trace over the zero modes yields the generic form \cite{periwal}: 
\begin{equation}
{\cal A} \sim  \delta (k_{1}^{0}+k_{2}^{0})\sqrt{\frac{1}{{\rm log}(q_{1})}}
f(q_{2},k_{2}^{0})  \label{result}
\end{equation}
The $\delta (k_{1}^{0}+k_{2}^{0})$
function arises from integrating over the zero modes of $\sigma $ model
fields $X^{0}(z,{\bar{z}})$ in a standard fashion. The amplitude (\ref
{anulus}, \ref{result}) is pathological, in the sense that it is divergent
as $q_{1}\rightarrow 0$,
 and requires regularization. 
It is this regularization which modifies the relation (\ref{result})
in order to ensure momentum conservation in the presence of string
loop effects. Seen from a scaffolder point of view, such effects
correspond to the recoil of the soliton in the scattering process. 

The pathological behaviour in the limit $q_{1}\rightarrow 0$ corresponds to
the pinched-anulus configuration: 
\begin{equation}
{\cal A}\sim g_{st}\int_{q\sim 0}\frac{dq_{1}}{q_{1}\sqrt{8\pi {\rm log}
(q_{1})}}A_{disk}(k_{1},k_{2})  \label{result2}
\end{equation}
where $A_{disc}(k_{1},k_{2})=<V(k_{1})V(k_{2})V^{i}V^{i}>$ is the tree-level
disc amplitude, with $\{V_{i}\}$ denoting a complete set of 
string states. 

One may cancel these infinities by adding to the action the so called {\em 
recoil} operators. These operators where introduced in 
\cite{recoil,kogmav,kogwheat,periwal} to
describe the back reaction (the recoil) of the soliton when two tachyons
scatter against it.
\pr
In this paper we are not interested in the role of the recoil operators from
the point of view of scattering in the scaffolding space--time {\em
per se}, but rather in the r\^oles of their couplings as generating quantum 
operators
of a phase space. However we find it useful for the convenience of the
reader to review the basic properties of these operators and their
origin. We do this in the next subsection and in the appendix. Further
details can be found in \cite{kogmav,kogwheat,periwal,diffusion}.

\subsection{D-branes and logarithmic operators}

As argued in refs. \cite{kogmav,periwal,kogwheat}, in order to
cancel the one-loop infrared divergence (\ref{result2}) one must add the
following tree-level closed-string (bulk) operator counterterm 
\begin{equation}
\delta {\cal A}=\int d^{2}z\partial _{\alpha }(f(X^{0})\partial ^{\alpha
}X^{i})  \label{oper}
\end{equation}
which contributes on the boundary. Its form is determined by general
properties of soliton backgrounds in string theory~\cite{kogmav}, and the
function $f(X^{0})$ in (\ref{oper}) is determined by requiring that the
above operator reproduce the infinities of the anulus amplitude (\ref
{result2}). The result is \cite{periwal}:
\begin{equation}
{V}_{imp}\equiv \int d^{2}z\,\partial _{\alpha }([u_{i}X^{0}]\Theta
(X^{0})\partial _{\alpha }X^{i})=\int d\tau \,u_{i}\left( X^{0}\Theta
(X^{0})\right) \partial _{n}X^{i})~;\qquad i=1,\ldots ,9.
\label{wrongrecoil}
\end{equation}
The step-function operator in (\ref{wrongrecoil}) needs to be defined
properly, and we adopt the integral representation~\cite{kogwheat}: 
\begin{equation}
\Theta _{\epsilon }(X^{0})=-i\int\limits_{-\infty }^{\infty }\frac{dq}{
q-i\epsilon }e^{iqX^{0}}\quad ,\qquad \epsilon \rightarrow 0^{+}
\label{theta}
\end{equation}
where $\epsilon $ is a 
regulator parameter. 
{}From the scaffolder point of view, 
the coupling constant $u_i$ may be considered as the 
change in the velocity 
of the brane due to recoil \cite{periwal,kogwheat}. 
This may be obtained by imposing overall conformal invariance of
the anulus and disc amplitudes~\cite{recoil}. 
As we shall discuss 
below, the cancellation of tree and
anulus divergences in our case is formally different 
from the approach of ref. \cite{recoil}. 
However, the basic qualitative features remain the same, 
and the above cancellation 
requires \cite{recoil,periwal}
\begin{equation}
u_{i}=8\sqrt{2}\pi g_{s}(k_{1}+k_{2})_{i}  \label{velocity}
\end{equation}
which expresses momentum conservation. This interpretation of
(\ref{velocity})
 is consistent with the fact that the soliton mass is proportional to $
1/g_{s}$, confirming the interpretation of $u_{i}$ as the D-brane velocity.
\pr
But this is not the end of the story.  Analysis of
the operator product of the operator (\ref{wrongrecoil}) reveals that it is
a {\it logarithmic } operator~\cite{gurarie} 
\begin{equation}
V_{imp}(\tau)V_{imp}(0)\sim {{\rm log}\tau\over\tau^2}  \label{log}
\end{equation}
Logarithmic operators, however, being related to degenerate solutions of
hypergeometric equations,
are known to come in {\it pairs} $C$ and $D$
\cite{gurarie,tsvelik}. We discuss the appearance and the properties
of such logarithmic operator
pairs in the appendix. 

To identify the pair in our case let us concentrate, for convenience,
on the $X^0$-dependent parts. It is clear that $X^0\Theta (X^0)$
appearing in (\ref {wrongrecoil}) plays the role of the logarithmic
$D$ operator~\cite{kogwheat}. Using the integral representation
(\ref{theta}) one may write it in the form
\begin{equation}
D_{\epsilon} = X^0\Theta _\epsilon (X^0) =
-i\int\limits_{-\infty}^\infty \frac{dq}{q - i\epsilon}X^0
e^{iq X^0} = -\int\limits_{-\infty}^\infty \frac{dq}{(q - i\epsilon)^2}
e^{iq X^0}  \label{deps}
\end{equation}
where we have integrated by parts. 

To find the corresponding $C$ operator we study the operator product of
$D_\epsilon$ 
(\ref{deps}) with the stress-energy tensor of the $\sigma$-model. 
Using the fact that the conformal dimension of the operator $e^{iq
X^0}$ is $q ^2/2$ one gets \cite{kogwheat} 
\begin{equation}
T(w)D_{\epsilon}(z)= -~\frac{\epsilon^2/2}{(w-z)^2} D_{\epsilon} + \frac{1}{
(w-z)^2}\epsilon\Theta(X^0)  \label{opetd1}
\end{equation}
{}From this, and making use of 
general properties of logarithmic operators, listed in the 
Appendix, we may identify the $C$ operator to be
\be 
C_{\epsilon} =  \epsilon \Theta_{\epsilon} (X^0)
\label{cop}
\ee
Note the factor of $\epsilon$. The action of the stress tensor on the
operators $C_{\epsilon}$ and $D_{\epsilon}$ reveals~\cite{kogwheat} 
that they have the same conformal dimension $\Delta=-{\frac{\epsilon^2}{2}}$
which is negative and vanishes in the limit $\epsilon \rightarrow 0^+$ (note
that this implies that the total dimension of the impulse operator (\ref
{wrongrecoil}), including the $\partial_{n}X^i$ factor, is 
$1-{\frac{\epsilon^2}{2}}$).  
We see that for non-zero $\epsilon$ the impulse operator
is {\em relevant} in a renormalization-group sense, and the resulting
theory is slightly non critical \cite{kogwheat,diffusion}.  

It is clear from (\ref{opetd1}) that we cannot work just with the $
D_{\epsilon}(X^0)$ operator because $C_{\epsilon}(X^0)$ will necessarily  be
induced by a scale transformation.  Thus, the proper recoil operator is
described by 
\begin{eqnarray}
{V}_{rec} = \int d\tau ~ [y_i C_{\epsilon}(X^0) \partial _nX^i +
u_iD_{\epsilon}(X^0) \partial _n X^i ]  \label{recoil}
\end{eqnarray}
where the coupling constants $y_i$ and $u_i$ in principle depend on
the world-sheet renormalization scale.

In ref. \cite{kogwheat} explicit expressions were derived 
for the one and two point functions 
of the logarithmic pair $C$ and $D$. For later use 
we list bellow the result for the two-point functions
(concentrating again for brevity on the $X^0$ parts):
\begin{eqnarray}
<C_\epsilon(z)C_\epsilon(0)> &\sim& -
\frac{1}{g_s}\epsilon^2\sqrt{\frac{\pi}{\alpha}}
\int\limits_{-\infty}^\infty \frac{dq}{(q^2+\epsilon^2)} e^{-2\eta
q^2\log|z/a|^2}  \nonumber \\
&\stackrel{\epsilon\to 0}{\sim}& \frac{1}{g_s}(0+O(\epsilon^2))  \label{cc3}
\end{eqnarray}
with $\alpha \equiv {\rm Log}|L/a|^2$, and $L$ and $a$ denoting the
world--sheet
infrared and ultraviolet cutoffs respectively, and
$\eta=1\, (-1)$ for Euclidean
(Minkowskian) signature of $X^0$ in the target space time of the 
scaffolder. 
Above we have re-instated the explicit powers 
of the string coupling constant, $g_s$, appropriate for 
the disc topology.

In a similar manner one has:
\begin{eqnarray}
<C_\epsilon(z)D_\epsilon(0)> &\sim& -\frac{\epsilon}{2g_s}\sqrt{\frac{\pi}{
\alpha}} \frac{\partial}{\partial \epsilon} \int\limits_{-\infty}^\infty 
\frac{dq}{q^2 + \epsilon ^2} e^{-2\eta q^2 \log|z/a|^2}  \nonumber \\
&\stackrel{\epsilon\to 0}{\sim}& {\frac{\pi}{2g_s}}\sqrt{{\frac{\pi}{
\epsilon^2\alpha}}} \left(1-2\eta\epsilon^2 \log|z/a|^2\right)  \label{cd}
\end{eqnarray}

Finally the two-point function for $D_\epsilon$ is given by 
\begin{eqnarray}
<D_\epsilon(z)D_\epsilon(0)> &=&\frac{1}{\epsilon^2} <C_\epsilon(z)
D_\epsilon(0)>  \nonumber \\
&\stackrel{\epsilon\to 0}{\sim}& {\frac{\pi}{2g_s}}\sqrt{{\frac{\pi}{
\epsilon^2\alpha}}} \left({\frac{1}{\epsilon^2}}-2\eta\log|z/a|^2\right)
\label{dd}
\end{eqnarray}
Performing the limit: 
\begin{equation}
\epsilon \rightarrow 0, \qquad \epsilon ^2 \log|L/a|^2 \sim O(1)
\label{epsilonlog}
\end{equation}
we obtain the canonical two-point correlation functions  
of logarithmic operators (c.f. Appendix). Notice however that $<DD>$
is singular in the limit.

Because the exact value of the numerical constant in (\ref{epsilonlog}) is a
free parameter we may choose it at will  (the difference between different
choices can be reabsorbed in the redefinition of the $\log z$ term) and thus
establish an unambiguous relation between $\epsilon$, the regularization
parameter in a target-space, and $L/a$, which is a world-sheet scale.
Choosing as normalization: 
\be
{\frac{1}{\epsilon^2}} =  \eta \log|L/a|^2 
\label{normal}
\ee
and then we get (up to a normalization factor) to zero$^{th}$ order
in $\epsilon^2$: 
\begin{eqnarray}
<C_{\epsilon}(z) C_{\epsilon}(0) > &\sim& 0  \nonumber \\
<C_{\epsilon}(z) D_{\epsilon}(0) > &\sim & {1\over 2 g_s}  \nonumber \\
<D_{\epsilon}(z)D_{\epsilon}(0)> &\sim&  {1\over 2 g_s}
\left(\log\left|{L\over a}\right|^2
- 2 \eta \log\left|{z\over a}\right|^2\right)  
\label{CD}
\end{eqnarray}
In the following, as we will consider the euclidean case (for the
scaffolding space-time), we set $\eta=1$.

To recapitulate, therefore, the requirement of 
cancelling anulus infinities in the scaffolder $\sigma$-model with 
matter (\ref{action}) one must include the pair of logarithmic
operators $C$ and $D$, which mix under scaling. 
The full action, which is free, therefore, from conformal anomalies
up to anulus level reads:
\bea
 S_\sigma &=& \int _{\Sigma} [\partial X^\mu {\overline \partial}X^\nu
\eta_{\mu\nu} + T(X)] + 
 \int d\tau [ x_i \partial _n X^i + v_i X^0
\partial_n X^i ] + \nn \\ 
&~&\int d\tau ~ [y_i C_{\epsilon}(X^0) \partial _nX^i +
u_iD_{\epsilon}(X^0) \partial _n X^i ]  
\label{action2}
\eea
with $D$ and $C$ given by (\ref{deps},\ref{cop}).

\section{The Space of Running Coupling Constants as a Phase Space}

In this and the following sections we will discuss how the couplings $y_i$ and 
$u_i$ of the operators $C$ and $D$ in (\ref{action2}) 
can be seen as the phase space of a point particle. 
{}From a $\sigma$-model view point 
we shall consider the action (\ref{action2}) 
with $T(X)=0$, $x_i=0$ and $v_i=0$. According to the 
discussion above this truncated $\sigma$-model is slightly
{\it non-critical}, and requires Liouville dressing for 
consistency. For our purposes we shall treat
the Liouville mode as independent from the target time $X^0$
of the scaffolder $\sigma$-model. This is due to the fact that
we are not interested in the scaffolder dynamics. 
The resulting Liouville-dressed couplings, then, 
$y_i$ and $u_i$ exhibit non-trivial world-sheet scaling. 
We shall study their behaviour 
under Liouville renormalization-group flow. 
Following general 
properties of non-critical strings \cite{emn,kogan}, then, we shall 
use the renormalization-group flow as a variable 
defining temporal evolution in this phase space, in the 
sense of Galilean invariance. 

\subsection{Finite-Size Scaling as Galilean Invariance}

Let us consider a finite size scaling~\cite{kogwheat} 
\begin{equation}
L \rightarrow L^{\prime}= L e^{t}  \label{fsscaling}
\end{equation}
Because of the  relation between $\epsilon$
and $L$ this transformation will change $\epsilon$
\begin{equation}
\epsilon^2 \rightarrow {\epsilon'}^2 = \frac{\epsilon^2}{1 + 2
\epsilon^2 t}  \label{epsilontransform}
\end{equation}
(note that if $\epsilon$ is infinitesimally small then $\epsilon^{\prime}$
is also infinitesimally small for any finite $t$) and we can deduce from
the scale dependence of the correlation functions (\ref{CD}) that 
$C_{\epsilon}$ and $D_{\epsilon}$ transform as: 
\begin{eqnarray}
D_{\epsilon} &\rightarrow& D_{\epsilon^{\prime}} = D_{\epsilon} + t
C_{\epsilon}  \nonumber \\
C_{\epsilon} &\rightarrow& C_{\epsilon^{\prime}}= C_{\epsilon}
\label{scaletr}
\end{eqnarray}
{}From this transformation  one can see that the coupling constants in front
of $D_{\epsilon}$ and $C_{\epsilon}$ in the operator defined in (\ref{recoil}),
i.e. the velocities $u_i$ and spatial collective coordinates $y_i$ of the
brane, must transform like: 
\begin{equation}
u_i \rightarrow u_i~~,~~y_i \rightarrow y_i +  u_i t  \label{scale2}
\end{equation}
And this is a Galilei transformation for the $y_i$ and $u_i$'s.

We have a first important result, the scale transformation of the
two--dimensional theory is a symmetry in the space of coupling
constants which coincides with a Galilei transformation.

We can consider this scaling as a renormalization flow on the world
sheet of the open string. 
In fact from \eqn{normal} and \eqn{fsscaling} we see that a scale
transformation is equivalent to a rescaling of $\epsilon$.
The corresponding $\beta$ functions are 
\begin{eqnarray}
&~&\beta ^{y_i} = -\frac{\epsilon ^2}{2} y_i + u_i  \nonumber \\
&~&\beta ^{u_i} = -\frac{\epsilon ^2}{2}u_i  \label{betas}
\end{eqnarray}
where the first part corresponds to the anomalous scaling dimensions of the
operators $C$ and $D$~\cite{kogwheat}. {}From (\ref{betas}) it becomes clear
that the coupling of the $D$ (momentum) operator in (\ref{recoil}) is not
exactly marginal as would have been required by the Galilean transformation,
were $t$ to be identified as time. There are small ${\cal O}[\epsilon^2]$ 
anomalous dimension terms that spoil this marginality. This
can be remedied by redefining the string coupling constant $g_s$ so as to
absorb such terms. Indeed, it is the coupling \cite{diffusion}
\begin{equation}
u_i^R \equiv \frac{u_i}{\epsilon}  
\label{ur}
\end{equation}
which is exactly marginal, 
\be
    du_i^R/dt = 0
\label{bur}
\ee        
and therefore plays the r\^ole of a Galilean
velocity. This amounts to a re-normalization of $g_s$ such that 
\begin{equation}
g_s^R =\frac{g_s}{\epsilon}  \label{gsr}
\end{equation}
is $t$-independent. 

The corresponding `renormalization' of the position 
\be 
   y_i^R \equiv \frac{y_i}{\epsilon} 
\label{yr}
\ee
yields the Galilean equation for (renormalized) velocities:
\be
  dy_i^R /dt = u_i^R 
\label{byr}
\ee
It is this set of renormalized couplings $y_i^R, u_i^R$ that  
constitute our classical phase space, which may be quantized 
canonically as we discuss in the subsequent sections. 

\subsection{Helmholtz Conditions and Quantization of Coupling Constants
\label{Helmholtz}}

The above analysis implies that in the presence 
of the logarithmic operators $C$ and $D$ 
the D-brane $\sigma$-model exhibits marginal deviations
from conformal invariance. It is however known \cite{ddk} that any
non-critical string $\sigma$-model can become conformal by introducing
a Liouville mode. The latter may be viewed \cite{emn} as a local
renormalization scale on a curved world sheet. For our case, this will
promote the first order equation \eqn{byr} to a second order one with
respect to the Liouville zero mode. This kind of equation can be derived from
a Lagrangian provided a certain set of conditions due to Helmholtz is 
satisfied \cite{hojman,emnd}. This is necessary and sufficient for
canonical quantization.

In fact \cite{polkle,emn,schm}, the 
most generic renormalization group flow for a  
$\sigma$--model coupling $\lambda_i$, corresponding to a vertex
operator $V_i$, in Liouville string theory 
has the form of a friction equation of motion \cite{emninfl}:
\be
\ddot \lambda_i (\phi ) + Q \dot \lambda_i (\phi )=
-\beta ^i (\lambda ) =
G^{ij}\frac{\partial}{\partial \lambda_j}
C (\lambda )
\label{eikosidyo}
\ee
where the dot represents differentiation to Liouville zero mode
$\phi$, $\beta^i$ is the flat
world-sheet renormalization group $\beta$ function, and 
$Q^2=C[\lambda ]-25$ is the central-charge deficit.
The quantity $C(\lambda)$ is the Zamolodchikov $C$-function
\cite{zam}. This function interpolates among conformal field theories
according to the $C$-theorem, which for flat world sheets reads:
\be
{\partial C\over\partial t}=-\beta^iG_{ij}\beta^j \label{cteo} 
\ee
and $G^{ij}\sim<V_iV_j>$ is the metric in theory space. 

In equation \eqn{eikosidyo} we took into account 
the gradient flow property of 
the $\beta$-functions:
\be 
     \partial _i C[\lambda] = G_{ij}\beta ^j 
\label{gradient}
\ee
which is an  
off-shell corollary of the flat-world-sheet
$C$-theorem~\cite{zam}. 
Equation (\ref{eikosidyo}) is characteristic
of frictional motion in a potential $C(\lambda_i )$. As known in
general such equations cannot be cast in a Lagrangian form. But in the
non-critical-string case this is possible due to the non-trivial
metric $G_{ij}$. In fact, the situation is entirely similar to the case
of inflationary theories \cite{emninfl}. 

The conditions for the existence of an underlying Lagrangian $L$ 
\cite{hojman} whose equations of motion are equivalent (but not 
necessarily identical) to (\ref{eikosidyo}), are equivalent to the
existence of a non-singular matrix $w_{ij}$ :
\be
w_{ij} ({\ddot \lambda}^j + Q {\dot \lambda}^j +\beta^j) =
\frac{d}{d t }(\frac{\partial L}{\partial {\dot \lambda}_i}   )-
\frac{\partial L}{\partial \lambda_i}
\label{triantadyo}
\ee
which obeys the following Helmholtz conditions :
\bea
w_{ij} &=& w_{ji} \nn \\
\frac{\partial w_{ij}}{\partial {\dot \lambda}^k} &=&
\frac{\partial w_{ik}}{\partial {\dot \lambda}^j}
\nn \\
   \frac{1}{2}\frac{D}{D \phi }
(w_{ik}\frac{\partial
f^k}{\partial {\dot \lambda}^j}
- w_{jk}\frac{\partial
f^k}{\partial {\dot \lambda}_i})&=&
w_{ik} \frac{\partial f^k}{\partial
\lambda ^j} -
w_{jk} \frac{\partial f^k}{\partial
\lambda_i}\nn \\
\frac{D}{D \phi} w_{ij} &=&
-\frac{1}{2} w_{ik}\frac{\partial
f^k}{\partial {\dot \lambda}^j}
-\frac{1}{2} w_{jk}\frac{\partial
f^k}{\partial {\dot \lambda}_i}
\label{33}
\eea
where
\be
f^i \equiv -Q {\dot \lambda }^i - \beta ^i(\lambda ) 
\qquad ; \qquad
\frac{D}{D \phi } \equiv \partial _t + {\dot \lambda^i} \partial _i
+ f^i \frac{\partial }{\partial {\dot \lambda}^i }
\label{34}
\ee
If the conditions (\ref{33}) are met, then
\be
  w_{ij}  =  \frac{\partial ^2 L } {\partial {\dot \lambda ^i}
\partial {\dot \lambda ^j}}
\label{35}
\ee
The Lagrangian in (\ref{35}) can be determined up to
total derivatives according to \cite{hojman}:
\bea
    {\cal S}  \equiv \int d\phi ' L &=&
    - \int d\phi ' \int _{0}^1 d\kappa \lambda ^i
    E_i (\phi ', \kappa \lambda, \kappa {\dot \lambda},
\kappa {\ddot \lambda} )\nonumber\\
 E_i (\phi ',\lambda,{\dot \lambda},{\ddot \lambda})  &\equiv&
  w_{ij}( {\ddot \lambda}^j
 + Q{\dot \lambda}^i + \beta ^i )
\label{36b}
\eea
In the case of non-critical strings one can identify \cite{emninfl}
$w_{ij}=-G_{ij}$. Near a fixed point, where the variation of $Q$ is small,
the action ${\cal S}$ becomes \cite{emninfl,emnd}:
\be
{\cal S}
= \int d\phi' ( -\frac{1}{2}{\dot \lambda}^i G_{ij}(\lambda,\phi)
  {\dot \lambda}^j
 -C[\lambda] + \dots )
\label{36c}
\ee
with the $\dots$ denoting
terms that can be removed by a renormalization scheme change.

Within a critical-string (on-shell) approach,
the action (\ref{36b},\ref{36c}) can be
considered as an effective
action generating the string scattering amplitudes.
Here it should be considered
as a target-space `off-shell' action
for non-critical strings~\cite{emn,emninfl}.
{}From \eqn{36c} it follows that the 
canonical momenta $p_i$ are 
given by:
\be
   p_i = G_{ij} {\dot \lambda}^j 
\label{momenta}
\ee

Let us now check the validity of the 
conditions (\ref{33}). 
We know that $G_{ij}$ is symmetric, so the
first of the Helmholtz conditions (\ref{33})
is satisfied. The next two conditions hold
automatically because of the gradient flow
property (\ref{gradient}) 
of the $\beta$ function,
and the fact that
$G_{ij}$ and $C[\lambda]$
are functions of the coordinates $\lambda ^i$
and not of the conjugate momenta. Finally,
the fourth Helmholtz condition provides the
condition
\be
  \frac{D}{D \phi } G_{ij} = Q G_{ij}
\label{37}
\ee
which implies an expanding scale factor for the metric in theory
space (`inflation')
\be
    G_{ij} [\phi,
    \lambda (\phi) ]= e^{Q\phi} {\hat G}_{ij} [\phi,
    \lambda (\phi) ]
\label{38}
\ee
where ${\hat G}_{ij}$ is a Liouville-renormalization-group invariant
function. This is exactly the form of the Zamolodchikov metric in 
Liouville strings \cite{emninfl,emnd}. Thus there is indeed
an underlying Lagrangian in the non-critical (Liouville) string
problem.

This allows canonical quantization, which as we will see in the next
section, is induced by higher genus effects \cite{emn,emnd}:
\be
[ \lambda ^i, \lambda ^j ] = 0
\qquad ; \qquad  [ p_i, p_j ] =0 \qquad ; \qquad
[\lambda ^i, p_j ] = -i \hbar \delta _j^i
\label{triantaena}
\ee
with the commutators being defined in theory space
and $p_i$ denoting the momentum \eqn{momenta} conjugate to $\lambda ^i$. 

Let us now come to the specific problem of $D$ branes. 
The pertinent system of flat-space 
Renormalization-Group flow equations is given 
by (\ref{bur},\ref{byr}). 
Liouville dressing is necessary for the position coupling 
$y_i^R$ only, since $u_i^R$ is {\it exactly} marginal. 
The corresponding central-charge deficit may be found by 
applying the flat world-sheet Zamolodchikov $C$-theorem \eqn{cteo}
for the deformation pertaining to the $y_i^R$ coupling: 
\be
      Q^2[y_i^R,u_i^R]  \propto  
C[t,y_i^R,u_i^R]-25 
\sim -\int _0^t dt' \frac{dy_i^R}{dt'} \epsilon ^2 G_{CC} \frac{dy_i^R}{dt'}
\sim -\frac{1}{g_s^R} \epsilon ^3 t u_i^Ru^{i,R} 
\label{deficit}
\ee
where $t$ is the flat world-sheet 
scale (\ref{normal}) and we used (\ref{cc3}).
As we discussed above, $t \epsilon ^2 \sim O[1]$, 
hence the central-charge deficit becomes of order 
$-\frac{\epsilon}{g_s^R}u_i^R{u^R}^i$.  
This choice corresponds to the identification of 
the logarithm of the size of the world-sheet disc 
with a (covariant) scale, which is nothing other than the 
zero mode of the Liouville field $\phi$.   
However, this identification {\it must} be made only at 
the very end of the computations, and hence 
at all intermediate steps the Liouville scale
should be treated independent of $\epsilon $. 
Moreover, the exactly marginal 
coupling $u_i^R$ is treated, of course, as {\it independent}
of $d{y_i}^R/dt$. 
It is then straightforward to see  
that the corresponding Liouville problem
satisfies the Helmholtz conditions 
(\ref{33}) with $f_i=-{\dot y}^{R}_i\sqrt{{\epsilon\over g^R_s} 
u_j^Ru^{jR}} - u_i^{R}$.

Therefore, in our case equation \eqn{eikosidyo} reads:
\be
{\ddot y}^{iR}+\left(\sqrt{{\epsilon\over g_s^R}
u_j^Ru^{jR}}\right)\dot y^{iR}=0 \label{inutile}
\ee
Near the fixed point $\epsilon\to 0$ this becomes the equation of a
free particle. Thus, the system of $\sigma$-model couplings  $y_i^R$
can be {\it quantized canonically} \'a la (\ref{triantaena}). 
Once we have established a canonical quantization we can then
take the flat world-sheet limit, provided we make the 
above identification of the world-sheet renormalization scale with the area 
of the disc. This will be assumed in what follows. 

The effective action (\ref{36c}), then,  
simply corresponds
to a free phase-space action for a particle of mass $\propto 1/g_s^R$:
\be
  {\cal S} = -\int dt \frac{1+ O[\epsilon]}{2g_s^R}u_i^Ru^{i,R}
\label{action3} 
\ee
consistent with the non-interacting brane background at 
the level of the scaffolder $\sigma$-model.
The fact that the mass of the particle is proportional to the 
inverse coupling constant of the string suggests that 
there is a natural length scale $\alpha'_s$ in the problem
\be
    \alpha _s' =(g_s^R)^2 \alpha '
\label{scale}
\ee
where $\alpha '$ is the  string Regge slope. Such a rescaling 
is characteristic of soliton strings \cite{shenker,dbranes,emnd}.

We now note that from the point of view of a D-particle (scaffolder) 
its mass is determined by
(\ref{velocity}). However in our construction, where the scaffolder 
is only an {\it auxiliary} entity used simply to induce 
the relevant deformations $C$ and $D$ in a formal manner,
there is {\it no} such {\it restriction}.
We can therefore just take $p_i= \kappa {u_i}^R$ with $\kappa$ an 
arbitrary scale with dimensions of mass.

Then, the 
mass of the particle whose action is described by (\ref{action3})
is
\be
     M = g_s^R\kappa^2 
\label{masspart}
\ee
As we
shall show in section 4 this point of view can be supported
explicitly by a $\sigma$-model path integral analysis 
on resummed (pinched) world sheets. 

An immediate question arises as to how one can 
define explicitly the 
$\hbar$ in theory space. Although the above approach 
tells us that a canonical quantization may be achieved,
however, it does not specify the details of the passage from 
classical to quantum mechanics. 
In string theory this can be achieved 
by a summation over genera on the world-sheet~\cite{emninfl}.
Below we shall review briefly the basic steps
by concentrating on the specific case of 
D-branes.  
As we shall argue, $\hbar$ will turn out to 
be proportional to the square of the string coupling constant. 

\section{$\sigma$-model Approach to the Quantization of Phase Space
\label{statistical}}

The necessity to go beyond genus zero 
world-sheet surfaces appears 
quite natural in
the framework of the scaffolder $\sigma$-model.
Since the effects associated with a {\it change} in a
conformal field theory background, such as recoil, or 
in general back-reaction of matter on the structure of
space-time etc, 
are purely stringy, the most natural way of
incorporating them in a $\sigma$-model language is to go {\it beyond a fixed
genus} $\sigma$-model and consider the effects of resummation over
world-sheet genera~\cite{emn,kogmav,recoil}. This 
is a very difficult procedure to be carried out
analytically. However, for our purposes a sufficient analysis, which
should describe the situation satisfactorily (at least at a
qualitative level), is
that of a heavy extended object in target space, which can be treated
semi-classically. From a first-quantized point of view, this implies
resumming one-loop (anulus) world-sheets. Non-trivial effects arise from
degenerate Riemann surfaces, namely from long-thin world-sheet strips
or tubes (for the closed-string case) 
(wormholes) that are attached to a Riemann surface of lower genus (disc in
our case).  

\subsection{Phase Space and Pinched World Sheets}  

We represent the Riemann surface $\Sigma$
as a strip connecting  two Riemann surfaces $\Sigma_1$ and $\Sigma_2$
of lower genus. We take account of the degenerate strip 
by inserting a complete set
of intermediate string states ${\cal E}_\alpha$ \cite{polchinski}.  Then
a generic $N$-point correlation function of vertex operators $V_i$
\begin{eqnarray}
A_N =\int dm_{\Sigma} \langle \prod_{i} \int d^2\xi_i V_{i}(\xi_i) \otimes
(ghosts) \rangle _{\Sigma } 
\end{eqnarray}
is given by the following expression 
\begin{eqnarray}
A_N \sim \sum _\alpha \int dq  \int d^2 z_1 ~\int d^2 z_2 \int
dm_{\Sigma_1\oplus\Sigma_2}  \nonumber \\
\langle \prod_{i} \int d^2\xi_i V_{i}(\xi_i) {\cal E}_\alpha (z_1) \otimes
q^{L_{0} -1}  \otimes {\cal E}_\alpha
(z_2) \otimes (ghosts) \rangle_{\Sigma_1 \oplus \Sigma_2}  \label{props}
\end{eqnarray}
where $q$ is the (real) modular parameter of the strip,
$\int dm$ denotes integration over moduli space, and ${\cal E}_\alpha$ are
the complete set of the intermediate states  with dimensions
$h_\alpha$ propagating along the thin strip
connecting the world-sheet pieces $\Sigma_1$ and $\Sigma_2$ (in our
case $\Sigma_1 = \Sigma_2$). The
terms `ghosts' indicate appropriate insertions of ghost fields, which,
although necessary for the target space reparametrization, are of no
consequences here.

One easily observes that logarithmic divergences in 
(\ref{props}), may come from states with $h_\alpha =0$
\cite{polchinski},  
in which case one has the infrared divergence at small $q
\rightarrow 0$ integral $\int dq / q$.  A
trivial example of such an operator is the identity, which however {\it does}
not lead to non-trivial effects, since it carries {\it zero} measure in the
space of states, and hence it does not contribute to (\ref{props}). On the
other hand, if there are states that are separated by a {\it gap} from the
continuum of states, i.e.\ {\it discrete} in the space of states, then they
bear non-trivial contributions to the sum-over-states and lead to
divergences in the amplitude (\ref{props}) \cite{recoil,kogmav,emn,schm,emnd}.

The logarithmic states we discussed earlier are precisely these
states.
There are various types of divergences associated with such states, 
as discussed in ref. \cite{kogmav}. 
First of all, 
there are leading divergent 
terms in (\ref{props}) arising due to the mixing between the states
corresponding to the logarithmic operators $C$ and $D$: 
\begin{eqnarray}
\mbox{\rm Diverg[STRIP]} & \sim & g_s \int_{q\sim\delta \to 0} \frac{ dq }{q } 
\ln q \int d^2 z_1 D(z_1) \int d^2 z_2 C(z_2)   \nonumber \\
& \sim & g_s (\ln^2 \delta) \int d^2 z_1 D(z_1) \int d^2 z_2
C(z_2)
\label{divstrip}
\end{eqnarray}
This gives the leading singularity $\ln^2 \delta$.

The effects of a dilute gas of wormholes on the sphere are assumed 
to exponentiate this bilocal operator
and one can obtain a change in the world-sheet action~\cite{recoil,kogmav} 
\begin{equation}
\Delta S \sim g_s (\ln^2 \delta) \int d^2 z_1 D(z_1)
\int d^2 z_2 C(z_2)
\end{equation}
This bilocal term can be written as a
local world-sheet effective action term, if one employs the well-known trick
of wormhole calculus~\cite{wormholes} by writing 
\begin{equation}
e^{\Delta S ^{CD}} \propto 
\int d\gamma_C d\gamma_D 
\exp\left[-\frac{1}{4g_s{\rm ln}^2\delta}G^{mn}\gamma_m
\gamma_n + \gamma _C \int d^2 z C +
\gamma _D \int d^2 z D \right]
\label{wlocal}
\end{equation}
In this way the 
bilocal operator can be represented as a local deformation on the world-sheet
of the string but with deformation couplings  $\gamma _m$, where $m = C$ or $D$
and $G^{mn}$ is  an {\it off-diagonal} 
metric in coupling-constant 
space, necessary to reproduce the initial bilocal
operator. The exponentiation of thin handles had been assumed above.
The off-diagonal metric in theory space includes all the appropriate
{\it normalization} factors ${\cal N}_m$, for the zero mode states.
Such factors are essentially the inverse of the $CD$ two-point
function of the operators
$C$ and $D$ given in  (\ref{cd}), which is {\it finite}~\cite{kogwheat}.  

Besides the leading divergent terms in (\ref{props}) there will be terms
with $\ln \delta$ behaviour, corresponding  to 
\bea
\Delta S^{D} &\equiv&
g_s{\rm ln}\delta\int d^2 z_1 D(z_1) \int d^2 z_2 D(z_2)\nn \\
\Delta S^{C} &\equiv&
g_s{\rm ln}\delta \int d^2 z_1 C(z_1) 
\int d^2 z_2 C(z_2)
\eea

Assuming their
exponentiation, one may use {\it new} wormhole parameters $\alpha _C$
and $\alpha _D$ to represent such counterterms in the world-sheet action as
above. The only difference now is that the distribution functions for the
wormhole parameters associated with the subleading $\ln\delta$
divergences are ordinary Gaussian associated with {\it symmetric} 
metrics  $G_{CC}$,$G_{DD}$ 
in coupling-constant space: 
\begin{equation}
e^{\Delta S^{D} + \Delta S^{C} } = \int d\alpha _C d\alpha _D e^{-\frac{
\alpha _c^2G^{CC}}{\Gamma _C ^2}} e^{-\frac{\alpha _D^2G^{DD}}{\Gamma _D^2}} e^{-\alpha
_C  \int d^2 zC_\epsilon -\alpha _D \int d^2z 
D_\epsilon }  \label{gaussct}
\end{equation}
where the widths $\Gamma _C^2$, $\Gamma _D^2$ 
are given by:
\be
\Gamma_C ^2 \sim 4g_s~\ln\delta, \qquad \Gamma_D ^2 \sim
4g_s~\ln\delta  \label{width}
\ee
and the diagonal metrics elements contain now zero-mode-state 
normalization factors
which are given by the absolute value of the inverse of the $CC$ (\ref{cc3}) 
and $DD$ (\ref{dd}) two-point functions
\be 
\left|G^{CC}\right| \sim \frac{1}{{\cal N}_C^2} \sim \frac{\epsilon ^2}{g_s}
\qquad G^{DD} \sim \frac{1}{{\cal N}_D^2} \sim \frac{1}{2g_s\epsilon^2 }
\label{diagonal}
\ee

{}From this analysis it becomes clear that the above-described 
pinched-anulus-corrected string 
matrix elements are divergent for $\epsilon ^2 
\rightarrow  0^+$.
{}From a two-dimensional (world-sheet) view-point 
one should either cancel or absorb the divergences
in coupling constants. If all the divergences 
could be canceled, then the $\sigma$-model would be conformally invariant,
whilst absorption of the divergent contributions 
to $\sigma$-model couplings would result in departures from
criticality~\cite{emn}. 

\subsubsection{Leading Divergences}

First, let us concentrate on the term (\ref{wlocal}). 
Following ref.\ \cite{recoil}, we consider the
propagation of two matter (closed-string) tachyon states $V_{1,2}$
in the background
of (\ref{wlocal}) at tree level. In such a computation 
the effects of the $C$-operator are subleading and can be ignored.  
The result is:
\be
{\cal A}_{CD} \equiv <V_1V_2 e^{\Delta S ^{CD}}>_{*} \sim
\int d\gamma_C d\gamma_D \exp\left[
-\frac{1}{4g_s{\rm ln}^2\delta}\gamma_m G^{mn}\gamma _n 
+
\gamma _D \int d^2 z D + \dots \right] V_1 V_2 
\label{dsv1v2}
\ee
where the $<\dots >_*$ denote averages with respect to a 
conformal 
$\sigma$-model action on the disc, and $\dots$ 
represent subleading terms.

The scaling property (\ref{scaletr}) must be taken into account.  
Under scaling on the world-sheet the $C$ operator emerges from $D$
due to mixing (see Appendix), with 
a scale-dependent coefficient $u_Dt$. This will contribute
to the scaling infinities we are considering here.  

The $D$ operator insertion in (\ref{dsv1v2})
needs to be {\it normal-ordered} in the disc. Normal ordering here 
amounts to {\it subtracting} scaling infinities originating from 
divergent contributions of the composite operator $D$ as 
$\epsilon \rightarrow 0^+$. To determine these infinities we 
first note that due to the $X^i$ 
parts of the operators $D$, their one-point function 
on the disc, computed with respect to the free $\sigma$-model action,
may be written as 
\be
   <e^{-\int d\tau D}>_* \sim 
e^{\frac{1}{2}\int_{\partial \Sigma} \int _{\partial \Sigma} 
d\tau d\tau ' <D(\tau)D(\tau ')>_*} 
\label{two-point}
\ee
As a consequence of (\ref{dd}) and \eqn{scaletr}, there are leading 
(scaling) divergences for $\epsilon\rightarrow 0^+$, which 
behave as $\frac{1}{4g_s} t^2 \gamma _D^2 \epsilon ^2 \sim 
\frac{1}{2g_s} \gamma _D^2 t$, using the normalization (\ref{normal})
between $t$ and $\epsilon$~\cite{kogwheat}.
Hence, normal ordering of the $D$ operator 
amounts to adding a term $-\frac{1}{2g_s}t\gamma_D^2$
in the exponent of (\ref{dsv1v2}) in order to {\it subtract} such divergences.

Let us now introduce a complete set of states 
$|\phi _\alpha >$ in the two-point function of string matter on the disc:
\be
<V_1V_2>_* = \sum _{\alpha }  
{\cal N}_\alpha^2 
<V_1 |\phi\alpha >_* < \phi\alpha |V_2>_*  
\label{completeset}
\ee
where ${\cal N}_\alpha ^2$ is a 
normalization factor for states. 
Taking into account (\ref{diagonal}), and  
the discussion above, 
according to which the effects of the $C$ operator
are also included in the $D$ under scaling
(\ref{scaletr}) \cite{kogwheat}, 
one observes that the leading 
divergent contributions  in (\ref{completeset}) 
are of the form 
\be
<V_1V_2>_* \sim - g_s~t <V_1|C><C|V_2> + {\rm subleading~terms} 
\label{vc}
\ee 
We now remark that the operator $C$ plays the r\^ole of the 
Goldstone modes for translation of  
the collective coordinates $X^i$ of the scaffolder, 
and as such one may 
apply the corresponding Ward identity in the $\sigma$-model 
path-integral to represent the action of $C$ on a physical 
state by $-\partial / \partial X^i$~\cite{recoil,emnd}. 
Thus, to this order in the string-coupling constant, the result 
may exponentiate to yield 
\be
   <V_1|V_2>_* \sim {\rm exp}\left(-{1\over 2}g_st
(\frac{\partial}{\partial X^i})^2\right)
\label{expon}
\ee
where we used the on--shell condition $V_j\left({\partial^2 \over \partial
{X^i}^2} V_i\right)=0$.

Then, (\ref{dsv1v2}) 
may be expressed as 
\bea 
 {\cal A}_{CD} &\sim&
\int d\gamma_D d\gamma_C
\exp\left[-\frac{1}{4g_s{\rm ln}^2\delta}\gamma _mG^{mn}\gamma _n  
-\frac{1}{2g_s}t \gamma _D^2 +  t 
\gamma_D^i \frac{\partial}{\partial X^i}
 - {1\over 2}g_st
\frac{\partial}{\partial X^i}  
\frac{\partial}{\partial X^i} + \dots \right]V_1V_2 
\nn \\
&=&\int d\gamma_D d\gamma _C\exp\left[-\frac{1}{4g_s{\rm ln}^2\delta}\gamma_m
G^{mn} 
\gamma_n 
-\frac{t}{2}( \frac{\gamma_D^i}{\sqrt{g_s}} 
- \sqrt{g_s}\frac{\partial}{\partial X^i})^2 +\dots \right] V_1V_2 
\label{cdsq2}
\eea
with the $\dots$ denoting finite terms. 
{}From (\ref{cdsq2}) it is evident that 
if one takes the limit $t \rightarrow \infty$ before the 
$\gamma _{D,C}$ integration, then  one obtains a 
$\delta $-function factor 
\be
      \delta ^{(d)}\left(\frac{(\gamma _D)_i}{\sqrt{g_s}} - 
\sqrt{g_s}(k_1 + k_2)_i\right)
\label{dfunc}
\ee
where $d$ denotes  the spatial dimension of the scaffolder $\sigma$-model, 
and $(k_{1,2})_{i}$ are the momenta of the matter closed string states.
{}From (\ref{dfunc}) one obtains 
\begin{equation}
u_i \sim g_s(k_1 + k_2)_i  
\label{velocityoptimum}
\end{equation}
This formula was mentioned earlier (\ref{velocity}),  
and is consistent with the $BPS$ mass of the stringy soliton $M_s
\propto \frac{1}{g_s}$~\cite{recoil}.
The above procedure guarantees  conformal invariance of the
$\sigma$-model, 
as far as leading divergences are concerned.
 This is the same result 
as the one obtained in ref. \cite{recoil}, but the procedure followed here 
is formally different.

For our purposes in this
work we shall reverse the logic, and consider a $\sigma$-model action
(\ref{action2}) with a tree-level coupling $u_i$ given by
(\ref{velocityoptimum}), but {\it opposite in sign}. The above analysis,
then, shows that resumming pinched genera in such a $\sigma$-model, leads
-in the limit $\epsilon \to 0$ - to a {\it cancellation} of the classical
(tree-level)  coupling constant $u_i$. As we shall show next, the
remaining subleading divergences yield {\it fluctuations} of the
phase-space couplings around their classical values.  This is the basis of
our construction of a quantum phase space, and it will be assumed in what
follows.

\subsubsection{Subleading Divergences}

We now proceed to treat the subleading divergences
of $O[\sqrt{ln\delta}]$, associated with (\ref{gaussct}).  
To this order the partition function of the $\sigma$-model 
on the resummed pinched topologies (RPT) assumes the form: 
\begin{eqnarray}
Z&=&\int DX d\alpha _m \exp [-\{S^* + y_i \int _{\partial \Sigma } 
\epsilon \Theta_\epsilon (X^0)\partial
_n X^i + u_i \int _{\partial \Sigma} X^0 \partial _n X^i \Theta
_\epsilon (X^0)
+  \nonumber \\
&~&\alpha _C \sqrt{\ln\delta} \int_{\partial \Sigma} 
\Theta(X^0)\partial _n X^i +
\alpha _D \sqrt{\ln\delta} \int _{\partial \Sigma} X^0\Theta(X^0)
\partial_n X^i + {1\over 4g_s}
G^{mn}\alpha_m\alpha _n \}]  \label{model}
\end{eqnarray}
where $S^*$ is the standard free $\sigma$-model action on the `bulk' of the
world sheet\footnote{Note that the normalization of $\alpha_C$ and $\alpha_D$
has been changed by $\sqrt{\ln\delta}$ as compared to \eqn{gaussct},
to facilitate the comparison with the tree-level result.}. 

{}From (\ref{model}) it, then, becomes clear that the effect of the operators 
$C$ and $D$ is simply to induce quantum fluctuations on $y_i$ and $u_i$.
Indeed by redefining 
\be
y_i + \alpha _C \sqrt{\ln\delta} \rightarrow {
\hat y}_i , \qquad  
u_i + \alpha _D \sqrt{\ln\delta} \rightarrow {\hat u}_i
\label{redef}
\ee 
and taking into account that the momentum $p_i$ of the $0$ brane is 
$\propto u_i/g_s$, we can rewrite
(\ref{model}) as a `quantum' target phase-space path integral : 
\begin{equation}
Z \sim \int d{\hat y}_i d{\hat p}_i \frac{1}{g_s \ln\delta} 
e^{-\frac{({\hat q}_m -q_m)G^{mn}({\hat q}_n - q_n)}{g_s \ln\delta}} 
\int DX
e^{-S^* -\int {\hat q}_m V_m(X)}  \label{model2}
\end{equation}
where $q_m \equiv \{ y_i, g_sp_i \}$, $V_m \equiv \{ C, D \} $. 

We first examine  
the Gaussian exponential factors in (\ref
{model2}). As we shall argue these are  
phase-space distribution functions for the 
{\it fluctuating} collective coordinates
and momenta of the $0$-brane. 
Thus the effect of resumming (pinched)
world-sheet surfaces is to induce a statistical gaussian spread of the
couplings. This is the essence of our {\it quantization} procedure.

To see this let us first concentrate on the velocity distribution
function  
\be
e^{-\frac{({\hat u}_i - u_i )^2}{g_s^2\epsilon ^2 
{\rm ln}\delta}}\equiv e^{-{(\delta u_i^{(q)})^2\over\Gamma_D^2}}
\label{velo}
\ee
with
\be
\Gamma _D^2 \sim (g_s^R)^2 \epsilon^2 {\rm ln}\delta 
\label{dwidth}
\ee
and $\delta u_i^{(q)}\equiv{\hat u_i-u_i\over\epsilon}$ denoting the
quantum fluctuations of the velocity. We
absorbed the $\epsilon^2$ dependences 
in a renormalization of the velocity operators, and hence of 
$g_s$ (see eqns.~\eqn{ur} and \eqn{gsr}). 
This way the fluctuations of the velocity maintain
their {\em exactly marginal} character. Indeed,  
from (\ref{model2}) it 
becomes evident that the renormalization (\ref{gsr}), 
sought for in the previous section, is a consequence of the summation
over pinched handles. Indeed, by 
making the 
identification  
of the infinities due to $\ln\delta$, $\delta \to 0$, with 
world-sheet scaling infinities~\cite{fs}, 
we observe that the $\alpha^i _D/\epsilon $
couplings may be marginal if we require that $\alpha _D$ behaves like the
classical $u_i$ under renormalization, i.e. 
\begin{equation}
\frac{d}{dt}\alpha ^i_D = -\frac{1}{2t}\alpha ^i_D  \label{ad}
\end{equation}
from which 
\begin{equation}
\delta u^{(q)}_i \equiv {\alpha_i}_D /\epsilon~:~ \beta_i^{\delta u} \equiv 
\frac{d}{dt} \delta u_i^{q} =0  \label{uq}
\end{equation}
may be considered as an {\it exactly marginal} fluctuation of (quantum)
velocity operators.

Repeating the reasoning above for
the distribution of the $C$ operator we find that its width $\Gamma_C^2$
behaves as 
\be
       \Gamma _C^2 \sim  (g_s^R)^2 \frac{1}{\epsilon ^2}{\rm ln}\delta 
\qquad t \rightarrow \infty 
\label{cwidth}
\ee
where, as before, the fluctuations for the position are defined by
$\delta y^{i,(q)} \equiv \frac{\hat y_i-y_i}{\epsilon}$.

In the same spirit,  
by requiring that ${\alpha_C}_i$ behaves like the
classical coordinate coupling $y_i$, 
\begin{equation}
\frac{d}{dt}\alpha^i_C = -\frac{1}{2t}\alpha ^i_C + \alpha ^i_D  \label{ac}
\end{equation}
one observes that the quantity:  
\begin{equation}
\delta y^{i,(q)} \equiv \alpha ^i_C/\epsilon~:~
\beta ^{\delta y} \equiv 
\frac{d}{dt}(\delta y^{i,(q)})=\delta u^{i,(q)}  \label{dy}
\end{equation}
plays the r\^ole of a position quantum fluctuation. 

It is those (renormalized) quantities
$\delta y^{(q)}$ and $\delta u^{(q)}$ that satisfy the canonical commutation
relations (\ref{triantaena}) in theory space, from which the
quantization of phase space
will emerge consistently, as we show in the next section. 
For the moment we note that a closer 
look at the widths (\ref{dwidth},\ref{cwidth})
of the gaussian fluctuations associated with the above quantities 
reveals an `{\it uncertainty}' relation: 
\be
   \Gamma ^2_D \Gamma ^2 _C  \sim g_R^2 {\rm ln}^2\delta 
\label{uncwidths}
\ee
{}From this relation, read as a Heisenberg uncertainty relation, one
can infer an upper bound in the Planck's constant in our phase space,
proportional to $g_R^2$. At the level we are presently working, it
suffices to view the right hand side of \eqn{uncwidths} 
of order of the square of Planck's constant.
Taking into account that the operator $D$ refers to velocities
and not momenta in target space, one may, then, infer from 
(\ref{uncwidths}) a minimum  uncertainty in phase space  
associated with Gaussian fluctuations,
with 
\be 
     \hbar _R \sim g_R^2 {\kappa}~{\rm ln}\delta  
\label{unc}
\ee
where the mass scale $\kappa$ was introduced in \eqn{unc},
and arises from the interpretation of $u_i$ as 
a velocity operator in a phase space with momenta 
$p_i \equiv \kappa~u_i$.
Recalling (\ref{masspart}) 
one observes that the particle mass $M = \kappa^2g_s$ 
is then {\it independent} of the scaffolder
soliton mass $1/(g_s\sqrt{\alpha '})$. 
We also observe that the subleading infinities associated with thin 
strips may be {\it absorbed} in a redefinition of the string scale 
$\alpha '_s$ (\ref{scale}) 
in order for the right-hand-side of (\ref{unc})
to be scale independent. 
With this in mind one then obtains a {\it definition} of Planck's constant 
in this problem proportional to the 
(string-loop-renormalized) string coupling constant\footnote{We 
remark that
in the case where $X^0$ is identified with the Liouville zero-mode $\phi$ 
\cite{emnd,diffusion} one obtains a similar expression for $\hbar$.
The physical interpretation, however, is different. One has a
{\it stochastic dynamics} 
for the phase-space evolution \cite{diffusion}. In such case the width
associated with the distribution of the $D$ operator 
(\ref{dwidth}) is still {\it scale independent}, whilst 
that of the $C$ operator  acquires a stochastic form 
$\Gamma _C ^2 \sim g_{R,\delta}^2 \phi$
with $g_{R,\delta} \equiv g_R {\rm ln}\delta $. The associated 
distribution
function may be {\it interpreted} \cite{diffusion} as 
a Gaussian {\it wave-packet} of a {\it free} D particle, 
spreading in Euclidean target time $\phi$ 
(i.e. diffusion)
with spread $(\hbar/m) \phi$ with $\hbar/m$ as in (\ref{unc}).}. 

\subsection{Phase-Space Dynamics}

Finally we come to the 
integral over the $X$ fields in (\ref{model2}), which 
will yield the target-space 
effective action of the scaffolder, depending 
on the (quantum) backgrounds ${\hat q}_i$. 
In the case at hand this will yield the {\it dynamics}
in phase space.  
There are two ways one can proceed, which 
depend on whether one identifies the renormalization scale 
with the target time $X^0$ of the scaffolder or not. 
In this latter case, the $X$ integration yields an effective 
action of a non-critical string theory. According to the 
analysis in ref. \cite{mm,emn,emnd} this is nothing other
than the Zamolodchikov $C$ function for the specific
problem. Using the $C$-theorem~\cite{zam}, then, one 
has:
\be
   S_{eff} \sim C[{\hat y}^R,{\hat u}^R, t] \sim 
-\int _0^t dt' \epsilon ^2 \beta^{y} G_{CC} \beta ^{y} =
-\int _0^t dt' \frac{1}{g_R}\epsilon  (u_i^R)^2 
\label{c-the}
\ee
due to (\ref{cc3}),(\ref{gsr}),(\ref{bur}),(\ref{byr}). 
This is the action of a free particle moving with constant 
velocity $u_i^R$, and thus the situation is in agreement 
with (\ref{action3}).

For completeness let us now compare this result 
with the one 
obtained under the identification of $X^0$ with 
the Liouville scale $\phi$, which is not necessary in this paper.  
In such a  case the scaffolder dynamics must be taken into 
account \cite{emnd,diffusion}.
The $X$ integration may then 
be performed in a standard
fashion in D-brane formalism, by noting the equivalence of the D-brane
background to that of an open string propagating 
in the background of imaginary gauge potentials
$A_M$ \cite{dbranes}. This yields 
the Born-Infeld action, 
$\int dX^0d^9X^i\sqrt{1-\left|F_{MN}^2(A)\right|}$, 
as the effective 
action obtained from $X$ integration. 

In our case, 
the particular form of the action (\ref{model},\ref{model2}), and 
eqs. (\ref{ur},\ref{bur})
suggest 
that the pertinent gauge potential is that of a space-constant 
`electric' field: 
\be
A_i = -i\epsilon\left({\hat y}_i + {\hat u}_i^R X^0\right)
\Theta_\epsilon (X^0)
\label{gauge}
\ee
However, unlike the standard D-brane result~\cite{dbranes}, 
here 
the electric field is singular at $X^0=0$~\cite{diffusion}:
\be
  iF_{0i} \equiv iE = \epsilon {\hat u}_i^R \Theta _\epsilon (X^0) + 
\epsilon\left({\hat y}_i + {\hat u}_i^R X^0\right)
\delta_\epsilon (X^0)
\label{field}
\ee

Were it not for the 
$\delta(X^0)$ term, the situation would be similar to 
the free non-relativistic particle discussed in this work. 
Indeed, from a Liouville renormalization point of view,
such infinities may be interpreted as infrared on the world-sheet,
since they correspond to the case where the size of the world sheet 
has shrunk to the order of the ultraviolet cut-off.  
One might, then, think of subtracting the $X^0=0$ 
(scaling) infinities from (\ref{field})
by some regularization procedure. Such terms
are not relevant for large-size world sheets as the ones used
in the present work.  
Then, the
pertinent effective action reads:
\be
   \int DX
e^{-S^* -\int {\hat q}_m V_m(X)} =e^{-S_{eff}[{\hat q}]}
\quad;\quad S_{eff} \propto \frac{1}{g_s}\int _{X^0>0} dX^0 d^9X^i\sqrt{1 - 
\epsilon^2 u_i^Ru^{i,R}}
\label{free}
\ee
and this would lead to the existence of 
an {\it upper bound} in velocities of order $\frac{1}{\epsilon}$. 
This bound  
is a consequence of the existence
of a critical field in the Born-Infeld Lagrangian~\cite{dbranes}.
In our case the critical velocity appears to be infinite as $\epsilon
\rightarrow 0^+$. This is consistent with the non-relativistic
nature of the slowly-moving $D$ brane we are employing here.
Upon expanding $S_{eff}$ in (\ref{free})
in powers of $\epsilon ^2$, and keeping  
the non-trivial 
leading terms as $\epsilon \rightarrow 0^+$
one obtains (for $X^0 >0$) the action of 
a free particle moving with constant velocity $u_i^R$, 
consistent with the result 
(\ref{action3}). 

Unfortunately the above procedure breaks down 
near $X^0=0$, i.e. for very small world sheets, 
which have to be taken into account if one identifies
the Liouville scale $\phi$ with $X^0$.  
In such cases 
the Born-Infeld Lagrangian exhibits non-trivial
mixing between phase-space
coordinates
${\hat y}$ and $u^R$, inducing singular interactions at $X^0=0$, 
which have been argued in ref. \cite{diffusion} 
to be responsible 
for decoherence. This makes an important 
difference from the treatment in this work, where the Liouville 
field is taken independent from the scaffolder time $X^0$.

\section{Commutators and Uncertainty Principle}

According to ref. \cite{emninfl} and the discussion in 
sub-section \ref{Helmholtz}, summation over world-sheet sheet genera
implies a {\it canonical} quantization in theory
space between the couplings $\{ \lambda^i \}$ and the conjugate momenta
defined in (\ref{momenta}): 
\begin{equation}
[\lambda ^i, G_{mj}\beta ^j ] = -i\hbar \delta ^i_m  \label{canonical}
\end{equation}
{}From the Helmholtz conditions \eqn{33} we know that 
the quantity $\hbar$ is just a constant in the sense that it
does not depend on $\lambda$ or the canonical momenta. It may depend
on the renormalization-group scale.

The operator
prescription of ref.\ \cite{emninfl} requires the replacement of classical
couplings $\lambda ^i$ by quantum operators in the expressions for 
$G_{ij}$ and $\beta ^i$, and the quantum ordering prescription
followed is the one that
leads to a stochastic quantum diffusion equation~\cite{emn,gisin}. 
Fortunately, for the $G_{ij}$ and to the order of $\epsilon$ we are 
considering, operator ordering issues will not arise.

In the D-brane $\sigma$--model, perturbed by the boundary recoil operators
(\ref{recoil}), the couplings are $y_i$ and $u_i =g_sp_i$, and the
non-vanishing Zamolodchikov metric components are $G_{CC} \sim <CC>$, 
$G_{DD} \sim <DD>$ and $G_{DC}=G_{CD} \sim <CD>$, given to leading order by
the two-point functions (\ref{cc3}),(\ref{dd}) and (\ref{cd}) respectively.
Higher-order corrections, in a (perturbative) expansion  over 
$\sigma$-model coupling constants, are incorporated in our approach 
by considering vacuum expectation values 
$< \dots >$ with respect to
an effective  
$\sigma$-model on the disc, deformed by the operators $C$ and $D$, but 
with the respective couplings representing 
the {\it quantum fluctuations} 
$\delta {\hat y}_i = {\hat y}_i -y_i$, 
$\delta {\hat p}_i \equiv {\hat p}_i-p_i$
around the classical phase space. As mentioned in section 4,
this follows from the requirement of 
cancellation of the leading divergences
in the pinched-world-sheet approximation. 

In the D-brane case, the pertinent operators are given by the 
fluctuations \eqn{uq} and \eqn{dy}. 
For the commutator of the $y_i$ coupling we have to leading
order in $\epsilon ^2$: 
\begin{equation}
-i\hbar \delta ^i_m = [\delta y^{i,q}, G_{CC}\beta ^{\delta y^m}]= \frac{
\epsilon ^2}{2} [{\hat y}_i, {\hat p}_m]  \label{ycom}
\end{equation}
{}From which
\begin{equation}
[{\hat y}_i, {\hat p}_m ]=-i{2\hbar\over\epsilon^2}\equiv\hbar_R 
\qquad~d\hbar_R/dt
=0  \label{ycomr}
\end{equation}
where $\hbar_R$ is defined in (\ref{unc})
and is {\em exactly marginal}. 
The necessity for the marginal character of $\hbar_R$ 
can be seen by deriving equation
\eqn{ycom} with respect to $t$, and taking \eqn{uq} and \eqn{dy} into
account. The constant $\hbar_R$  plays the r\^ole of
Planck's constant in our phase space.

To discover stringy effects, one should evaluate $G_{CC}$ not in the 
free action but in the presence of $C$ and $D$ deformations.
In a perturbative expansion in the couplings of the above deformation,
one can 
bring down one more power
of the couplings. 
An analysis similar to the one in ref. \cite{kogwheat}, then, shows 
that
the dominant contributions come from the $D$
deformations, and are of order 
$-<\!CCD\!>g_s^R\epsilon ^2\delta {\hat p}_i \sim
-\epsilon ^2 (\delta {\hat p}_i)$, where $\delta {\hat p}_i = 
{\hat p}_i - p_i $ denotes the quantum fluctuations around a 
classical momentum. The sub-dominant $C$ perturbation may also be easily
incorporated by adding a term $-<\!CCC\!> \epsilon \delta {\hat y}
\sim {\epsilon^3\over g^R_s}
 (\delta {\hat y})$
in the $G_{CC}$ metric. Thus, proceeding as above, the next-to-leading order
coupling-constant corrections to the commutator (\ref{ycomr}) read: 
\begin{equation}
[{\hat y}_i, {\hat p}_m ] \sim -i\hbar_R \delta^i_m (1 - 
{\epsilon \over g_s^R}\delta 
{\hat y}_m - 2 g_s^R (\delta {\hat p}_m) + \dots )^{-1}  \label{ycomcorr}
\end{equation}

We now notice that the presence of $(\delta {\hat p})$- and $(\delta 
{\hat y})$- dependent denominators in the commutators (\ref{ycomcorr}) implies
stringy corrections to the uncertainty. To this end let us first take into
account the Gaussian form (\ref{model2}),(\ref{gaussct}) of the distribution
functions of the fluctuations $\delta {\hat y}$ and $\delta {\hat p}$, which
implies that any average of linear functions of the respective commutators
vanish.

Thus, from a general property of self-adjoint operators, such as $\hat y$
and ${\hat p}$, one can derive from (\ref{ycomcorr}) 
the resulting uncertainty: 
\begin{equation}
\Delta y \Delta p \ge \frac{\hbar_R}{2} \left| <\frac{1}{1 -
{\epsilon \over g_s^R}
\delta{\hat y}
-2 g_s^R \delta{\hat p}+ \dots }>\right|
\sim \frac{\hbar_R}{2} (1 + {\epsilon ^2\over (g_s^R)^2} \Delta y^2 + 
O(\alpha_s^{\prime})\Delta p ^2 + \dots )  \label{uncert3}
\end{equation}
reinstating dimensions with respect to the new D-brane scale (\ref{scale})
$\alpha _s ^{\prime}=
(g_s^R)^2 \alpha ^{\prime}$, depending on the string coupling constant. For
weakly coupled strings $g _s^R << 1$ this is smaller than the standard Regge
scale. In the limit $\epsilon \rightarrow 0$ the uncertainty (\ref{uncert3})
is of the form of the string uncertainty principle \cite{ven}.

This result is in agreement with the Heisenberg-Microscope approach to
the uncertainty principle for D-branes obtained in \cite{kogwheat}.
We also remark that in the context of non-critical strings analogous
uncertainty relations lead to measurability bounds for distances,
consistent with quantum decoherence \cite{amelino,diffusion}.

\section{Conclusions and Discussion}

In this paper we have argued that the couplings of the vertex
operators describing interaction of a D-brane with string matter,
can give rise to a quantum phase space. We did
so in the simplest possible case, the lowest dimensional soliton (a
0-brane) in a flat background and ignoring supersymmetry. 
Moreover we resummed only pinched topologies of the world-sheet.  
We found that the quantum particle described this way is a free
nonrelativistic particle. Less simple solitons and backgrounds should
give rise to other (more realistic) structures.

We only
considered single-brane backgrounds. Multiple D-branes and possible bound
states among them or among strings and D-branes have been ignored. Such
structures can also be represented in a conformally invariant way, as argued in
ref. \cite{bound}, at least in flat target-space backgrounds. 

In \cite{bound}
the case of $n$ parallel Dirichlet $p$-branes has been considered in a flat
target space. 
We
recall~\cite{dbranes} that the D-brane operator (\ref{dbraneop}) or (\ref
{recoil}) describes transverse excitations of the brane. The collective
coordinates $Y_i(X^0, \dots X^p)$ may be viewed~\cite{bound,callan} as
scalar fields in the world-volume of the brane $\{ X^0, \dots X^p\}$. On the
other hand, the longitudinal excitations 
\begin{equation}
A_I (X^0, \dots X^p) \partial _\tau X^I \qquad I=0,\dots , p  \label{long}
\end{equation}
are associated with gauge fields on this world volume. For a single brane
the gauge group is $U(1)$. According to the analysis of ref. \cite{bound},
$n$ identical $p$-branes on top of each
other are described by an effective
low-energy field theory on the world volume, with a world-volume 
$U(n)$ gauge field and $9-p$ scalar fields in the adjoint
representation, representing the collective coordinates $X^i$.
The effective field theory on
the world volume of the D-brane is defined, then, as the dimensional
reduction of the ten-dimensional supersymmetric Yang-Mills field theory with
gauge group $U(n)$. The bosonic part of the effective scalar potential of
such a theory is given by 
\begin{equation}
V=\frac{T^2}{2}\sum _{i,j=p+1}^{9} Tr[X^i,X^j]^2  \label{potential}
\end{equation}
where $T \propto (\alpha ^{\prime})^{-1}$ is the D-brane tension, and the
trace is over the adjoint representation of $U(n)$. 

In the above
description the configuration space of the `collective coordinates' of the
system of identical $n$ D-branes consists of effectively {\it
noncommuting} matrices. When the D-branes are far apart, a classical vacuum
solution, with unbroken supersymmetry, minimizing the potential $V=0$ in (\ref
{potential}), is the dominant configuration, and one obtains a set of
{\it commuting } $X^i$. In that case the matrices $X^i$ can be
simultaneously 
diagonalized
\cite{bound}, with
each diagonal entry representing the target-space location of the $k$-th brane,
$k=1,\dots, n$. 

At short distances things are by no means simple.
When the branes are {\it very close} to each other, closer than the scale 
$\sqrt{\alpha _s ^{\prime}}$, massive excitations in the spectrum of the $U(n)
$ gauge theory must be taken into account. In ref. \cite{bound} it was
argued that in that case, one should consider the full quantum theory of the 
$U(n)$ field theory on the world volume, and it might be that states with
broken supersymmetry ($[X^i,X^j] \ne 0$) might arise. If this were the case,
this would imply that at {\it very short distances} the stringy space time
would have a {\it non-commutative} structure.

{}From the results of ref. \cite{emnd}, and the present work, then, one
might speculate that, in such a case, summing over world-sheet topologies of
the underlying  $\sigma$-model might lead to a proper {\it 
quantization} of a {\it noncommutative } space-time. Consequently,
following the spirit of refs. \cite{zumi,dopli,kempf}, one could arrive at
new forms of
uncertainty relations involving only the coordinates of space time, which
would be up and above the phase-space uncertainty relation (\ref{uncert3}).

At present there is no known example of such a situation within the
flat-space D-brane theory. We expect that curved target-space backgrounds
in string theory 
play an important r\^ole in such issues. We base this
expectation on our experience with two-dimensional $\sigma$-model
supersymmetric field theories~\cite{emn,yung}. Indeed, the $N=2$
supersymmetric $\sigma$ model on a $SL(2,R)/U(1)$ target manifold, which
from a target-space-time point of view describes physics near the
singularity of a black hole in string theory~\cite{witt,emn}, admits
world-sheet instantons in its spectrum~\cite{yung}. These are responsible for
a breaking of the world-sheet supersymmetry in the sense of a `false'
vacuum, characterized by a non-zero vacuum energy. The world-sheet conformal
invariance of the model is broken~\cite{yung,emn}. From a non-critical
string point of view, therefore, this situation arguably
describes (unstable) vacua of a quantum string theory which should be taken
into account. 

One can imagine a similar situation happening in our
world-volume theory: some non-trivial metric configurations for
the collective coordinates of the 
(quantum) D-branes produce a dynamical
breaking of supersymmetry, via some world-brane non-perturbative effects.
Probably, also in this case, non-criticality plays a r\^ole analogous to
the one played in the present paper, as a result of
the incompatibility of the Dirichlet boundary conditions with world-sheet
conformal invariance~\cite{dbranes,li}. 

Needless to say, much more work is needed before even tentative
conclusions are reached as regards the quantum structure of
space-time.
Nevertheless, we believe that the present work provides
compelling evidence for the 
importance of solitonic string backgrounds in the structure of
quantum physics and the structure of space time.

\section*{\Large {\bf Appendix: Logarithmic Operators}}

In this appendix we discuss the appearance of pairs of 
logarithmic operators in the
correlation functions of conformal field theories. We will limit
ourselves to a sketchy introduction of them in the theories of
interest to us. For more details we refer the interested reader 
to the literature, and in
particular to \cite{gurarie,tsvelik,flohr,kogmav,kogwheat}.

It is well known that in general, hypergeometic equations of the kind:
\begin{equation}
x(1-x)\frac{d^2 {\cal F}}{dx^2} + [c - (a+b+1)x] \frac{d{\cal F}}{dx} - ab 
{\cal F} = 0  \label{hypereq}
\end{equation}
admit two independent solutions (see for example \cite{ww,gr})
\begin{equation}
{\cal F}_1 = F(a,b,c; x), ~~~~~~~ {\cal F}_2 = x^{1-c} F(a-c+1,b-c+1,2-c; x)
\end{equation}
where $F(a,b,c; x)$ is the hypergeometric function.

However in the case in which $c$ is an integer the two solutions would
be degenerate, and they are, in the case of
$c>1$: 
\begin{eqnarray}
{\cal F}_1 &=& F(a,b,1+m; x),  \nonumber \\
{\cal F}_2 &=& \ln x~ F(a,b,1+m; x) + H(x),  \label{1+m}
\end{eqnarray}
where $H(x) = x^{-m}\sum_{k=0}^{\infty}h_k x^k$ and $h_{m} =0$, unless
either $a$ or $b$ equal $1+m^{\prime}$ with $m^{\prime}$ a natural
number $m^{\prime}<m$. In this case the second solution is only a
polynomial in $x^{-1}$ and no logarithmic terms are present. In case
of $c<1$ instead 
\begin{eqnarray}
{\cal F}_1 &=& x^m F(a+m,b+m,1+m; x),  \nonumber \\
{\cal F}_2 &=& \ln x~ x^m F(a+m,b+m,1+m; x) + H(x),  \label{1-m}
\end{eqnarray}
where $H(x)$ is again some regular expansion without logarithms, unless
either $a$ or $b$ equal $-m^{\prime}$ with an integer $m^{\prime}$ such that 
$0 \leq m^{\prime}<m$, in which case both solutions do not have logarithmic
terms also.

A similar phenomenon appears in the Operator Product Expansions (OPE) of
vertex operators. Equations of the type \eqn{hypereq} appear the
conformal blocks of the theory:
\begin{equation}
<V_1(z_1) V_2(z_2) V_2(z_3) V_1(z_4)> = \frac{1}{(z_1 - z_4)^{2h}(z_2 -
z_3)^{2h}} {\cal F}(x),  \label{F}
\end{equation}
where $x = (z_1 - z_2)(z_3 - z_4)/(z_1 - z_4)(z_3 - z_2)$  and only the
dependence on holomorphic coordinates $z_i$  has been written for
simplicity, and the $V$'s are primary fields with the {\it same} 
conformal dimension $h$.
 
In this case the logarithmic terms of \eqn{1+m} or \eqn{1-m} yield
an anomalous  OPE:
\begin{equation}
V_1(z_1)~ V_2(z_2) = ... + (z_1-z_2)^{h_C - 2h } (\bar{z}_1-\bar{z}_2)^{
\bar{h}_C - 2\bar{h} } \left[D + C \ln|z_1-z_2|^2 \right] + .... ,
\label{AB}
\end{equation}
where the dimension $h_C$ of the operators $C$ and $D$ is determined by the
leading logarithmic terms in the conformal block (\ref{F}).  We have 
written here both $z-$ and $\bar{z}-$dependences explicitly  and it is
important to note that the logarithmic term depends on $|z|$, even for
chiral fields, because in the full conformal blocks  actually $\ln |z|$
appears, as was shown in \cite{tsvelik}.

The OPE of the stress-energy  tensor $T$ with these
fields is: 
\begin{eqnarray}
T(z) C(0) &=& {\frac{h_C }{z^2}} C(0)+ {\frac{1}{z}} \partial_z C(0) + ... 
\nonumber \\
T(z) D(0) &=& {\frac{h_C }{z^2}} D(0)+{\frac{1}{z^2}} C(0)+{\frac{1}{z}}
\partial_z D(0) + ...  \label{JOPE}
\end{eqnarray}
for general logarithmic pair $C$ and $D$ with
anomalous dimensions $h$.  
We see immediately that we have a mixing between $C$
and $D$.  The Virasoro operator $L_0$ is not diagonal  and mixes $C$
and $D$ inside a $2\times 2$ Jordan cell 
\begin{equation}
L_0 |C> =h_C |C>; \qquad L_0|D>=h_C |D> + |C>  \label{oneb}
\end{equation}

Substituting (\ref{AB}) into the four-point correlation function
yields \cite{tsvelik}) the following  two-point
correlation functions for the fields $C$ and $D$: 
\begin{eqnarray}
\langle C(x) D(y)\rangle &=& \langle C(y) D(x) \rangle = \frac{\kappa}{
2~(x-y)^{2h_C }}  \nonumber \\
\langle D(x) D(y)\rangle &=& \frac{1}{(x-y)^{2h_C}} \left(-\kappa
\ln|x-y|^2 + d\right)  \label{CC} \\
\langle C(x) C(y)\rangle &=& 0  \nonumber
\end{eqnarray}
This is our defining property of the logarithmic pair.

Note that the constant $d$ is arbitrary and can be changed by shifting
$D \rightarrow D  + $
const $C$. The coefficient $\kappa$ is defined by the leading  logarithmic
term in the conformal block.

The $C-D$ mixing has consequences for the string propagator on the 
world-sheet cylinder
between states $|m>$ and $|n>$,  which is defined as 
\be
\int dq d{\overline q} <n|q^{L_0 -1} {\overline q}^ {{\overline L}_0 -1} |m>
\ee
where $q = \exp (2\pi i \tau)$ and $\tau$ is the modular parameter. In the
usual case, when $L_{0},~ {\overline L}_0$ are diagonal,  one gets after
integrating over $\tau$ 
\begin{eqnarray}
<n|\frac{1}{L_0 +{\overline L}_0 -2}~\delta(L_{0}- {\overline L}_0) |m>,
\end{eqnarray}
where $\delta(L_{0}- {\overline L}_0)$ enforce the condition $h = \bar{h}$
for all propagating states.  However in the case of logarithmic operators 
one must take into account the Jordan  cell structure of $L_0,~{\overline L}
_0$,  which in the sector of $|CD>$ and $|{\overline CD}>$  states leads to 
\begin{eqnarray}
q^{L_0} = q^{h_C} \left( 
\begin{array}{cc}
1 & \ln q \\ 
0 & 1
\end{array}
\right); ~~~~ q^{{\overline L}_0} = {\overline q}^{\bar{h}_C} \left( 
\begin{array}{cc}
1 & \ln {\overline q} \\ 
0 & 1
\end{array}
\right)
\end{eqnarray}

We thus have logarithmic factors appearing.
Note that $\ln q$ factors arise also in the characters 
$Tr_{h}q^{L_0}$  as was discussed by Flohr in \cite{flohr}.

For us the propagator takes the form:
\be
\int dq d{\overline q}q^{h_C -1} {\overline q}^ {\bar{h}_C -1} <CD|\left( 
\begin{array}{cc}
1 & \ln q \\ 
0 & 1
\end{array}
\right)|CD><{\overline C}{\overline D}|\left( 
\begin{array}{cc}
1 & \ln \bar{q} \\ 
0 & 1
\end{array}
\right)|{\overline C}{\overline D}>  \label{CDprop}
\ee
One has either $\ln q$ or $\ln \bar{q}$ terms for transitions when either $D$
goes to $C$ or $\bar{D}$ goes to $\bar{C}$ and $\ln q \ln \bar{q}  = 4\pi^2
|\tau|^2$ for transition $D\bar{D}$ to $C\bar{C}$.

In our case we are interested in modes with conformal dimension
$h_C \to 0$, in which case one obtains extra logarithmic infinities
as $q\to 0$. 

\nk {\large \bf Acknowledegments}

We wish to thank Giovanni Amelino-Camelia
for many illuminating conversations, and 
collaboration at an early stage of this work. 
We would also like to thank John Ellis, Ian Kogan, 
and John Wheater 
for numerous discussions
and comments. F.L. wishes to thank the Department 
of (Theoretical) Physics of Oxford University for hospitality.

\end{document}